\documentstyle[aps,multicol,epsf,epsfig]{revtex}
\begin{document}

\title{A Novel Predictive Tool in Nanoengineering: 
Straightforward Estimation of Superconformal Filling Efficiency}

\author{A. De Virgiliis, O. Azzaroni, 
R. C. Salvarezza and  E. V. Albano}

\address{Instituto de Investigaciones Fisicoqu\'{\i}micas Te\'{o}ricas
y Aplicadas, (INIFTA), CONICET, UNLP.
Sucursal 4, Casilla de Correo 16, (1900) La Plata. ARGENTINA. 
FAX : 0054-221-4254642.
E-mail : ealbano@inifta.unlp.edu.ar}

\date{\today}

\maketitle

\vskip 0.5 true cm

\begin{abstract}
It is shown that the superconformal filling (SCF) efficiency
($\epsilon_{SCF}$) of nano-scale cavities can be
rationalized in terms of relevant physical and geometric parameters.
Based on extensive numerical simulations and using the dynamic
scaling theory of interface growth, it is concluded that the
relevant quantity for the evaluation of $\epsilon_{SCF}$ is the
so-called {\bf physical} aspect ratio $S_{P} = L/M^{\beta/\alpha}$,
where $\alpha$ ($\beta$) is the roughness (growth) exponent that
governs the dynamic evolution of the system and
$L$ ($M$) is the typical depth (width) of the cavity.
The theoretical predictions are in excellent agreement with
recently reported experimental data for the SCF of electrodeposited
copper and chemically deposited silver in confined geometries,
thus giving the basis of a new tool to
manage nanoengineering-related problems not completely resolved so far.
\end{abstract}

\begin{multicols}{2}

Modern trends in technology require the development of straightforward 
routes to the fabrication of ultrasmall devices
with complex architectures \cite{Timp}. At present one of the 
most efficient procedures appears to be the filling of molds 
having nanometer sized patterns with a depositing material 
\cite{JESMoffat,Andri}. 
This method is a promising strategy for nanofabrication in several 
technological fields such as nanowire production, information storage,
nanoelectronics, nanoelectromechanical systems, etc.. A crucial point 
for the application of this strategy is that the complete
filling \cite{West,Taiwan} of nanometer sized complex cavities present in the mold,
i.e., the so-called superconformal filling (SCF), is required \cite{PRLMoffat}.
This is a non-trivial problem of increased complexity as the aspect ratio 
$S = L/M$ increases, where $L$ ($M$) is the typical depth (width) 
of the structure to be filled. Significant advances towards
highly efficient SCF have been achieved in both electrochemical 
deposition (ECD)\cite{SalvaPRL1} and chemical 
vapor deposition (CVD)\cite{SalvaPRL2,eisen}, which are 
techniques widely employed in serial fabrication of microdevices.
In contrast to the progress in the development of experimental 
and technical aspects of this subject, a theoretical framework
to understand the mechanisms involved in SCF is still lacking. 
This theoretical knowledge would be the starting point of new 
tools in nanoengineering design.
  
Recently, interface evolution upon both CVD and ECD growth, has
succesfuly been accounted for by the so-called Kardar-Parisi-Zhang (KPZ) 
growth mode \cite{kpz}. Thus, KPZ has been used for modeling  
conformal growth \cite{zangwill,kahanda} in complex architectures, 
i.e., a technique closely related to SCF. These results open the possibility 
of exploring the SCF problem with the aid of the framework already
established for the study of interface dynamics \cite{bara}.
In this letter we develop quantitative relationships between relevant
interface dynamic parameters, the aspect ratio of the architecture to
be filled, and the SCF efficiency. Thus, for a given dynamic behavior of
a growing system under nanoscale confinement, the SCF efficiency for
each architecture can be predicted. 

Metal particles are deposited in narrow cavities of depth $L$ and 
width $M$ (see figures 1(a-b)) according to both the KPZ \cite{kpz} 
and the EW \cite{EW} growth modes by means of Monte Carlo simulations
of a two-dimensional lattice model \cite{bunde} (three-dimensional simulations
are limited to small system sizes) \cite{SalvaPRL2}. The formation of a $bcc$
crystal is considered and the cavity dimensions are meassured in lattice units
(l.u.). So, the ranges $20 \leq L \leq 1600$, $20 \leq M \leq 1600$,
and $1 \leq S \leq 8$, where $S = L/M$ is the aspect ratio of the 
cavity, are considered. 

Figure 1 shows typical growth regimes obtained assuming the KPZ mechanism.
The deposition process starts at the surface of the cavity and the
development of a rough growing interface, which runs essentially parallel
to the substrates, can be observed, as expected for a conformal filling
process (figure 1(a)). During these early time stages, the width of the
interface, defined as the r.m.s. of average deposit height (figure 1(b)),
increases according to $W(t) \sim t^{\beta}$, where $\beta > 0$ is the 
growth exponent. As the system evolves the correlation length in the 
direction parallel to the growing interface increases as $\xi \sim t^{1/z}$,
where $z$ is the dynamic exponent. If the development of the interface is
not interrupted by the collision among growing interfaces
within the cavity, eventually the correlation length reaches
the size of the system, e.g. $\xi \approx L$ since the depth 
is the relevant length scale in this case. At this
long-time regime the interface width saturates at some constant
value $W_{sat} \sim L^{\alpha}$, where $\alpha$ is the
roughness exponent. This type of behavior is known as the
the Family-Vicsek scaling approach \cite{FV} that has proved
to be very successful for the description of the dynamic evolution
of growing interfaces, namely
\begin{equation}
W(L,t)\sim L^\alpha f(\frac t{L^z})  ,
\label{eq1}
\end{equation}
where $f(u)$ is a suitable scaling function that behaves as follows:
(i) $f(u)= constant$ for $u \gg 1$ so that the
interface width saturates for a long enough time and
(ii) $f(u)\sim u^\beta $ for $u\ll 1$. The
former condition implies that $W(t)\sim t^\beta $ holds during the
short-time regime. A scaling relationship can easily be derived
so that $z =\frac \alpha \beta$ and only two independent exponents
remain.

\begin{figure}
\narrowtext
\centerline{{\epsfxsize=8cm \epsffile{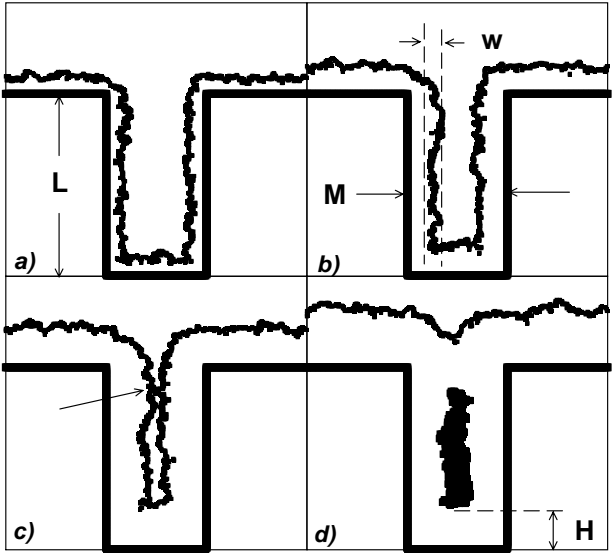}}}
\vskip .5 true cm
\caption{Typical configurations obtained upon cavity
filling according to the KPZ growth mode.
Black points show the location of the growing
interfaces. Relevant quantities such as the depth ($L$) and the
width ($M$) of the cavity, as well as the width ($W$) of the
interface and the distance from the lowest inaccessible site 
to the bottom of the cavity ($H$), are shown. $a)$, $b)$, $c)$ and $d)$ 
are obtained upon deposition of 15, 25, 35 and 45 monolayers, 
respectively. The arrow in $c)$ shows the location of the first
interface collision event. The black area in $d)$ correspond
to unfilled sites.}
\label{fig1}
\end{figure}

As the SCF process further develops, the topmost sites of 
the longest interfaces, running in the direction parallel
to $M$, eventually collide. Notice that figure 1(c) has been 
selected in order to show the first
collision event. Since deposition proceeds through mass
transport via the solution or the vapor phase, empty sites below
the collision point become inaccessible and can no longer be filled. 
Collision events and the consequent generation of inaccessible sites
are due to stochastic fluctuations of the interface width and take
place during the latest stages of the SCF process. Finally, a number
$N_{I}$ of inaccessible sites along the center of the cavity 
remains empty (figure 1(d)) limiting the quality of the deposit,
\cite{JESMoffat}.
Also at this stage, the depletion of the interface close to the 
center of the cavity is observed (figure 1(d)) as experimentally
observed \cite{West}. So, the SCF efficiency ($\epsilon_{SCF}$)
can be defined as

\begin{equation}
\epsilon_{SCF} = 1 - (N_{I}/LM)  
\label{eq2}
\end{equation}      

Figure 2(a) shows plots of $\epsilon_{SCF}$ {\it vs} $S$ obtained by
filling cavities of different size. Values of $\epsilon_{SCF}$ 
corresponding to various experiments, estimated after proper
digitalization of the published images, which have also been included
in figure 2(a), indicate an excellent qualitative agreement with
the simulation results. As follows from figure 2, the filling 
efficiency not only depends on the aspect ratio but also on the
dimensions of the cavity. Such a dependence can be derived using
equation (\ref{eq1}) and heuristic arguments.
In fact, at the final stage of deposition one has

\begin{equation}
N_{I} \approx 2W (M - H),
\label{eq3}
\end{equation}

\noindent where $H$ is the average heigth of the lowest
inaccessible site (see figure 1(c)). Furthermore, the height
of the deposit as measured from the bottom of the cavity is 
of the order of $M/2$, so that $H =  BM/2$, where $B$ is a 
constant close to unity. At this point one has to consider
two possible scenarios, namely: [{\it i}] ([{\it ii}]) the 
collision of the interfaces occurs after (before) the saturation 
of the interface width. So, defining $\chi \equiv 1 - \epsilon_{SCF}$ 
and using equation (\ref{eq1}) within the interface saturation 
regime, for the case [{\it i}] one gets 
  
\begin{equation}
\chi^{\it i} = 2AL^{\alpha} (L - BM/2)/LM = 
2\,A\,S\, L^{\alpha - 1}\,[1 - B/2S] ,
\label{eq4}
\end{equation}

\noindent where $A$ is a constant of the order of unity.

\begin{figure}
\narrowtext
\centerline{{\epsfxsize=9cm \epsffile{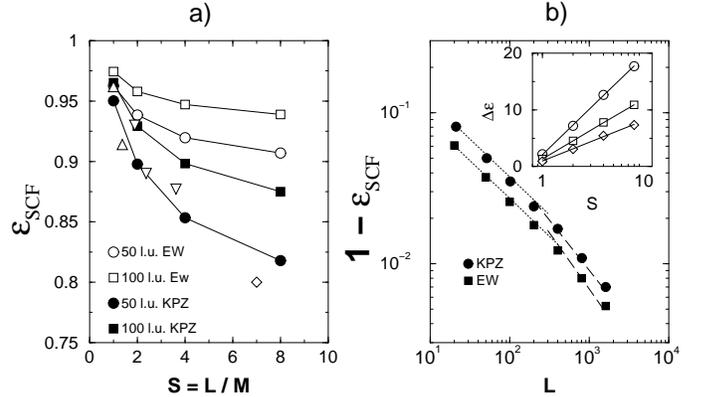}}}
\vskip .5 true cm
\caption{(a) Cavity filling efficiency $\epsilon_{SCF}$ $vs$
the aspect ratio $S$ obtained for cavities of different width $M$ as
listed in the figure and assuming both KPZ and EW growth modes.
$\bigtriangleup$, $\bigtriangledown$ and $\Diamond$ correspond to
experimental results on copper electrodeposition ([2] and [6]),
and on silver chemical vapor deposition [9], respectively.
(b) $\chi=1-\epsilon_{SCF}$ $vs$ the depth
of the cavity $L$, for $S=1$. The dotted lines have slopes $\alpha-1=-0.5$,
and dashed lines have slopes $\beta-1=-0.66$ (KPZ) and $\beta-1=-0.75$ (EW),
respectively. Inset: relative increase of the SCF efficiency
$\Delta\epsilon$ {\it vs} $S$ for different values of the width $M$.} 
\label{fig2}
\end{figure}

On the other hand, if the collision occurs before interface 
saturation (case [{\it ii}]), replacing  $W(t)\sim t_{coll}^\beta $
with a collision time $t_{coll} \sim M/2$ in equations (\ref{eq2})
and (\ref{eq3}), it follows that

\begin{equation}
\chi^{\it ii} = C\,M^{\beta - 1}\,[1 - B/2S] ,
\label{eq5}
\end{equation}

\noindent where $C$ is a constant of the order of unity.

Figure 2(b) shows log-log plots of $\chi$ {\it vs} $L$.
For the KPZ growth mode the crossover between 
two different regimes can clearly be observed at a certain 
crossover length $L_{c}$. In fact, 
for $L < L_{c} \approx 280 $, which corresponds to
the collision of saturated interfaces (case [{\it i}]), the obtained 
straight line has slope $\alpha-1=-0.5$ (see equation (\ref{eq4})), 
in excellent agreement with the well-known (exact) value 
$\alpha=1/2$ of the KPZ growth mode \cite{bara}. Furthermore, for $L > L_{c}$
the regime of non-saturated interface collision 
is observed and the straight line has slope $\beta-1=-0.66$
(see equation (\ref{eq5})), also in excelent agreement with the 
exact value $\beta=1/3$ for KPZ \cite{bara}. The same trend is 
also observed for the EW growth mode and the lines
corresponding to the different regimes have slopes 
$\alpha-1=-0.5$ and $\beta-1=-0.75$, respectively, also in excellent
agreement with $\alpha=1/2$ and $\beta=1/4$ expected for the EW model.

Figures 2(a) and (b) also show that for the EW process the values of
$\epsilon_{SCF}$ are higher than those of the KPZ growth mode.
This finding is in agreement with the well known fact that KPZ interfaces
are rougher than the EW ones. The inset in figure 2(b) shows, for different
values of $S$, the relative increase in $\epsilon_{SCF}$ for both growth
modes, defined as
$\Delta\epsilon\equiv(\epsilon^{EW}-\epsilon^{KPZ})/\epsilon^{KPZ}$. 
Furthermore, it is interesting to note that the growth
mode upon metal electrodeposition can be modified by placing
suitable additives into the plating bath. In this way unstable
interfaces turn into smoother ones \cite{SalvaPRL1,Gewirth}. For this reason 
the presence of additives has been found to be an essential requirement 
for the achievement of high SCF efficiency \cite{Taiwan}. It should be remarked 
that this scenario is fully consistent with our theoretical predictions.

At the crossover length the values of $\chi$ given by 
equations (\ref{eq4}) and (\ref{eq5}) must be the same, 
so it follows that 

\begin{equation}
L_{c} = D\,M^{\beta/\alpha} = D\, M^{1/z} ,
\label{eq6}
\end{equation}

\noindent where $D$ is a constant. This relationship is consistent
with the fact that the dynamic exponent $z$ sets the crossover
length scales upon self-affine interface growth \cite{bara}. Furthermore,
equation (\ref{eq6}) strongly suggests that the physical 
aspect ratio $S_{P} = L / M^{1/z}$ is the relevant magnitude
for the evaluation of the cavity filling efficiency, 
instead of the geometric one $S = L/M$. 
In fact, $S_{P}$ actually accounts for the two different regimes that 
may dominate SCF efficiency: for $S_{P} \ll 1$ the filling
efficiency is greater, but decreases upon increasing the
depth of the cavity, since the colliding interfaces 
have no longer fully developed their maximum roughness.
Furthermore, for $S_{P} \gg 1$ the filling eficiency is smaller
and decreases faster when the depth of the cavity is increased,
due to the fact that the colliding interfaces exhibit their
maximum roughness.

Finally it should be noted that we are dealing with a two-dimensional model.
Three-dimensional simulations are more realistic but are limited to small systems
sizes and experiment long crossovers that would turn unreliable the interpretation
of the interface dynamic parameters \cite{bunde}. It can be also argued that a
three-dimensional model would give different results; however, our model predictions
are in well agreement with data taken from cross-section images of systems grown in
three dimensions, as shown in figure 2(a).

In conclusion, based on a well-established scaling
approach for the description of growing interfaces, a
powerful predictive tool for a straightforward estimation
of the SCF efficiency has been developed. Only the knowledge
of the dynamic $z=\frac \alpha \beta$ exponent is required
to predict the ability of a given process to fill nanocavities. 
Therefore, all the experimental information obtained during the
last years on growing dynamics using planar substrates \cite{Zhaobook}
can be managed in order to predict the processing conditions of
complex nano-architectures.

{\bf  ACKNOWLEDGMENTS}. This work was financially supported by
CONICET, UNLP and ANPCyT (Argentina).

\vskip -0.5 true cm

\end{multicols}

\end{document}